\documentclass[aps,prl,twocolumn,superscriptaddress]{revtex4}
\pdfoutput=1
\usepackage{graphicx}
\usepackage{color}
\usepackage{amssymb}
\usepackage[colorlinks,urlcolor=blue,citecolor=blue]{hyperref}
\def\hhref#1{\href{http://arxiv.org/abs/#1}{arXiv:#1}} 

\begin{document}

\title{Time-Dependent Hartree-Fock Solution of Gross-Neveu  models:
Twisted Kink Constituents of Baryons and Breathers}
\author{Gerald V. Dunne}
\affiliation{Physics Department, University of Connecticut, Storrs CT 06269, USA}
\author{Michael Thies}
\affiliation{Institut f\"ur  Theoretische Physik, Universit\"at Erlangen-N\"urnberg, D-91058,
Erlangen, Germany }

\begin{abstract}

We find the general solution to the time-dependent Hartree-Fock problem for 
the Gross-Neveu models, with both discrete (GN$_2$) and continuous (NJL$_2$) chiral symmetry. We find new multi-baryon, multi-breather and twisted breather solutions, and 
show that all GN$_2$ baryons and breathers are composed of constituent twisted kinks of the NJL$_2$ model. 

\end{abstract}


                                          
\maketitle

Self-interacting fermion systems describe  a wide range of physical phenomena in particle,   condensed matter and atomic physics \cite{Nambu:1961tp,Fulde:1964zz,peierls,Heeger:1988zz,rajagopal,casalbuoni,zwierlein,pitaevskii,Adams:2012th,Herzog:2007ij}. Applications include solitons, excitons, polaritons, breathers and  inhomogeneous phases  in superconductors, conducting polymers, liquid crystals, particle physics, and cold atomic gases, and also illustrate the widespread  phenomenon of induced fermion number \cite{jackiw}.
The  Gross-Neveu models (${\rm GN}_2$ and ${\rm NJL}_2$)  in 1+1 dim. quantum field theory describe $N$ species of massless, 
self-interacting Dirac fermions
 \cite{Gross:1974jv}: 
\begin{eqnarray}
{\cal L}_{\rm GN} &=& \bar{\psi} i\partial \!\!\!/ \psi + \frac{g^2}{2} \left( \bar{\psi} \psi \right)^2
\label{eq:gn}\\
{\cal L}_{\rm NJL} &=&   \bar{\psi} i\partial \!\!\!/ \psi + \frac{g^2}{2} \left[\left(\bar{\psi} \psi \right)^2-\left(  \bar{\psi} i \gamma_5\psi \right)^2\right].
\label{eq:njl}
\end{eqnarray}
These are soluble paradigms of symmetry breaking phenomena in strong interaction particle physics and  condensed matter physics \cite{Nambu:1961tp,Thies:2006ti}. In the 't~Hooft limit, 
$N \to \infty$,  $Ng^2=$ constant,  semiclassical methods become exact, as pioneered in this context by Dashen, Hasslacher and Neveu (DHN) \cite{Dashen:1974ci,Feinberg:2003qz}. Classically,
the ${\rm GN}_2$ model has a discrete chiral symmetry, while the ${\rm NJL}_2$ model has a continuous chiral symmetry. At  finite temperature and density, and at large $N$,  these 
models exhibit inhomogeneous phases with crystalline condensates,  directly associated with chiral symmetry breaking  \cite{Basar:2009fg}. 
The basic physics of these GN phases is the Peierls effect of condensed matter physics \cite{peierls,horovitz,braz,machida}. 
This analysis of equilibrium thermodynamics is based on exact spatially inhomogeneous solutions to the gap equation, or equivalently the Hartree-Fock problem, which solves the Dirac 
equation subject to constraints on the scalar and pseudoscalar condensates \cite{Basar:2009fg,Takahashi:2012pk}. 
Here we extend these  results to  the complete exact solution of the {\it time-dependent} Hartree-Fock (TDHF) problem, relevant for  scattering processes, 
transport phenomena and non-equilibrium physics:
\begin{eqnarray}
&&{\rm GN}_2: (i \partial \!\!\!/ - S(x, t))\psi_{\alpha} = 0\, ; \,\, S=-g^2 \sum_{\beta}^{\rm occ} \bar{\psi}_{\beta} \psi_{\beta}
\label{eq:gn-tdhf}
 \\
&&{\rm NJL}_2 : (i \partial \!\!\!/ - S(x, t) -i \gamma_5 P(x, t))\psi_{\alpha} = 0\,\, ;
\nonumber \\
&& S=-g^2 \sum_{\beta}^{\rm occ} \bar{\psi}_{\beta} \psi_{\beta} \,\, ,
\,\, P=-g^2 \sum_{\beta}^{\rm occ} \bar{\psi}_{\beta}i \gamma_5 \psi_{\beta}
\label{eq:njl-tdhf}
\end{eqnarray}
We solve these TDHF problems in full generality, describing the dynamics, including scattering, of non-trivial topological objects such as kinks, baryons and breathers. 
Surprisingly, we  found that the most efficient strategy is to solve the (apparently more complicated) NJL$_2$ model first, and then obtain GN$_2$ solutions by imposing further constraints 
on these solutions. This reveals, for example, that the GN$_2$ baryons and breathers found by Dashen, Hasslacher and Neveu \cite{Dashen:1974ci} are in fact  bound objects of 
twisted NJL$_2$ kinks, and  that the scattering of  GN$_2$ baryons and breathers can be deduced from the scattering of twisted kinks.
This includes new breather and multi-breather solutions in NJL$_2$, as well as new multi-baryon and multi-breather solutions for GN$_2$.

We stress that while it is well known that the classical equations  for the GN$_2$ and NJL$_2$ models  are closely related to integrable models \cite{integrable}, this fact is only 
directly useful for the solution of the TDHF problem for the simplest case of kink scattering in the GN$_2$ model, which reduces to  the integrable Sinh-Gordon equation. The more general 
self-consistent TDHF solutions to (\ref{eq:gn-tdhf}, \ref{eq:njl-tdhf}) involving twisted kinks,  baryons and breathers, do not satisfy the Sinh-Gordon equation; instead we find a  
general ``master equation'' [see Eq. (\ref{eq:master})], whose solution 
reduces to a finite algebraic problem solvable in terms of determinants. 

We also stress that these more general solutions require a self-consistency condition relating the filling fraction of valence fermion states to the parameters of the condensate solution, 
as for the static GN$_2$ baryon \cite{Dashen:1974ci}, the static twisted kink \cite{Shei:1976mn}, and the GN$_2$ breather \cite{Dashen:1974ci}. For our time-dependent solutions, this 
important fact means that during scattering processes there is non-trivial back-reaction between fermions and their associated condensates and densities 
\cite{Dunne:2011wu}. Kink scattering in the GN$_2$  model,  described by Sinh-Gordon solitons \cite{Klotzek:2010gp,Jevicki:2009uz}, is much simpler as there is no  self-consistency condition 
or back-reaction.

With Dirac matrices, $\gamma^0 = \sigma_1,  \gamma^1 = i \sigma_2, \gamma_5 = - \sigma_3$, 
and light-cone coordinates (note $\bar z$ is {\it not} the complex conjugate of $z$): $
z=x-t,  \bar{z} = x+t,  \partial_0 = \bar{\partial}-\partial,  \partial_1 = \bar{\partial} + \partial 
$, the Dirac equation in (\ref{eq:njl-tdhf})  is:
\begin{equation}
2i \bar{\partial}\psi_2 = \Delta \psi_1\, ,  \quad 2i \partial \psi_1 = -\Delta^* \psi_2 \, , \quad \Delta \equiv S-iP
\label{eq:dirac}
\end{equation}
Write the complex potential $\Delta$ and   continuum spinor $\psi_\zeta$:
\begin{eqnarray}
\Delta  =  \frac{\cal N}{\cal D}\qquad, \qquad 
\psi_{\zeta} = \frac{e^{i\left( \zeta \bar{z}-z/\zeta\right)/2}}{{\cal D}\,\sqrt{1+\zeta^2}} \left( \begin{array}{c} \zeta{\cal N}_1 \\
- {\cal N}_2 \end{array} \right)
\label{eq:ansatz}
\end{eqnarray}
where ${\cal D}$ is real, and the complex light-cone spectral parameter $\zeta$ is related to the 
 energy $E$ and momentum $k$ as: 
$k = \frac{1}{2}\left( \zeta- \frac{1}{\zeta} \right)$, $E = - \frac{1}{2} \left( \zeta + \frac{1}{\zeta} \right)
$,
in units of $m$, the dynamically generated fermion mass.  The ansatz for $\psi_{\zeta}$ anticipates the fact that the potential $\Delta$ is
transparent.

We solve (\ref{eq:dirac}), and  associated TDHF consistency conditions, using an ansatz method, 
positing a  
decomposition with a finite number $n$ of simple poles:
\begin{equation}
{\cal N}_{1,2}(\zeta)  =  {\cal N}_{1,2}^{(0)} + \sum_{i=1}^n \frac{1}{\zeta-\zeta_i} {\cal N}_{1,2}^{(i)}
\label{eq:poles}
\end{equation}
Matching powers of $\zeta$ we learn that ${\cal D}$ and ${\cal N}$ must satisfy the ``master equation''
\begin{equation}
 4\,  \partial \bar{\partial}\ln {\cal D}=1-|\Delta|^2 
\label{eq:master}
\end{equation}
in addition to various sum rules obeyed by the residues ${\cal N}_{1,2}^{(i)}$ \cite{dt-inprep}. 
 Furthermore, the following equations must hold for all $i=1,\dots n$,
\begin{eqnarray}
 2i ( {\cal D} \bar{\partial}-\bar{\partial}{\cal D} ) {\cal N}_2^{(i)} - \zeta_i ( {\cal D} {\cal N}_2^{(i)} -   {\cal N} {\cal N}_1^{(i)} ) & = & 0
\nonumber \\
 2i \zeta_i  ( {\cal D} \partial  -  \partial {\cal D} ){\cal N}_1^{(i)}  +  {\cal D} {\cal N}_1^{(i)} -   {\cal N}^* {\cal N}_2^{(i)}  & = & 0
\label{eq:sumrule}
\end{eqnarray}
The residues of $\psi_{\zeta}$ at the poles $\zeta=\zeta_i$ provide normalizable bound state spinor solutions:
\begin{equation}
\psi^{(i)} =  \frac{1}{{\cal D} \, V_i} \left( \begin{array}{c} \zeta_i {\cal N}_1^{(i)} \\ - {\cal N}_2^{(i)} \end{array} \right)\, \, , \quad V_i  \equiv   e^{-i(\zeta_i \bar{z}-z/\zeta_i)/2} 
\label{eq:bound1}
\end{equation} 
An alternative set of normalizable bound state spinors comes from 
$\psi_\zeta$ at the 
complex 
conjugate poles $\zeta_i^*$:
\begin{equation}
\phi^{(i)} = \frac{V_i^*}{\cal D} \left( \begin{array}{c} {\zeta}_i^* {\cal N}_1({\zeta}_i^*) \\ -{\cal N}_2({\zeta}_i^*) \end{array} \right)
\label{eq:bound2}
\end{equation}
These two sets of bound states are linearly related,
$ \psi^{(i)} = \sum_j \Omega_{ij} \phi^{(j)}$. The condition that ${\cal D}$ is real and has no zeroes restricts the matrix $\Omega$ to the form
\begin{equation}
\Omega_{ij}=i\, \hat{\Omega}_{ij}/(\zeta_j^*)^2
\label{eq:Omegahat}
\end{equation}
where $\hat{\Omega}$ is a positive definite hermitean matrix.
Together with (\ref{eq:poles}) we obtain a finite dimensional algebraic system: 
\begin{eqnarray}
&&{\cal N}_1({\zeta}_j^*) - \sum_{i,k}\frac{1}{\zeta_i(-{\zeta}_j^*+\zeta_i)}V_i \Omega_{ik} V_k^* 
\tilde{\zeta}_k {\cal N}_1({\zeta}_k^*) =  {\cal D} 
\nonumber \\
&& {\cal N}_2({\zeta}_j^*) +  \sum_{i,k}\frac{1}{-{\zeta}_j^*+\zeta_i} V_i  \Omega_{ik} V_k^*
 {\cal N}_2({\zeta}_k^*) =  {\cal N} 
\label{eq:algebraic}
\end{eqnarray}
We have found a remarkably simple solution to this algebraic system, which yields a compact determinant expression for all the 
ansatz quantities in the TDHF solution:
\begin{eqnarray}
{\cal D}=\det(\omega+B)\,\, &,&\,\, {\cal N}=\det(\omega+A)\nonumber\\
 {\cal N}_1(\zeta)=\det(\omega+C)
\,\, &,& \,\, {\cal N}_2(\zeta)=\det(\omega+D)
\label{eq:dets}
\end{eqnarray}
with matrices 
\begin{eqnarray}
B_{ij}  & = &  \frac{i\, V_i^* V_j}{\zeta_j-{\zeta}_i^*} =  \frac{\zeta_i^*}{\zeta_j} A_{ij}\,\, ,
\nonumber \\
C_{ij} & = &    \frac{\zeta-{\zeta}_i^*}{\zeta-\zeta_j} B_{ij} = \frac{\zeta_i^*}{\zeta_j} D_{ij} \,\, ,
\label{eq:full}
\end{eqnarray}
where $\omega$ is a positive definite hermitean matrix: $\omega_{ij}=\zeta_i^* \hat{\Omega}^{-1}_{ij}\zeta_j$. 
This gives the complete solution to the Dirac equation for time-dependent  transparent (complex) potential $\Delta$, and via the master equation (\ref{eq:master}) gives  
a natural relativistic generalization of the 
Kay-Moses general transparent static  Schr\"odinger  potential \cite{kaymoses}
\begin{eqnarray}
V(x)  =   -  \partial_x^2  \ln  {\rm det} ({\bf 1}+ A) \, , \,\, 
A_{ij}  =  \sqrt{a_i a_j} \frac{e^{(\kappa_i+ \kappa_j)x}}{\kappa_i+\kappa_j}
\label{eq:kaymoses}
\end{eqnarray}
and its time-dependent Schr\"odinger  generalization \cite{nogami}. It also provides a new  closed-form solution to the finite algebraic problem, found recently in 
\cite{Takahashi:2012pk}, for the {\it static} transparent NJL$_2$ Dirac equation.

We now show that this solution also gives a self-consistent solution to the fully quantized TDHF problem (\ref{eq:njl-tdhf}), provided certain filling-fraction conditions 
are satisfied by the combined soliton-fermion system, generalizing the conditions already found by DHN, Jackiw-Rebbi, and Shei \cite{Dashen:1974ci,Shei:1976mn,jackiw}.
Consider first the induced fermion density in the Dirac sea. Introducing a cut-off scale $\Lambda$, \begin{eqnarray}
\rho_{\rm ind}  &=&  \int_{1/\Lambda}^{\Lambda}  \frac{d\zeta}{2\pi} \frac{\zeta^2+1}{2\zeta^2} \left( \psi_{\zeta}^{\dagger} \psi_{\zeta} -1 \right)\label{eq:sea}\\
&=& \int_{1/\Lambda}^{\Lambda} \frac{d\zeta}{2\pi} \frac{1}{2\zeta^2 {\cal D}^2} \left( \zeta^2 (|{\cal N}_1|^2-{\cal D}^2 ) + |{\cal N}_2|^2-
{\cal D}^2 \right)  
\nonumber
\end{eqnarray}
The pole ansatz (\ref{eq:poles}), a partial fraction decomposition, and the known asymptotic behavior of the ansatz functions, 
combine to show that the linear and logarithmic divergent terms cancel, 
leading to the finite result:
\begin{equation}
\rho_{\rm ind} =  \frac{i}{4\pi} \sum_{i,j} \frac{{\zeta}_i^*  \zeta_j{\cal N}_1^{(i)*}{\cal N}_1^{(j)} +
{\cal N}_2^{(i)*}{\cal N}_2^{(j)}}{{\cal D}^2\, V_i^*\,V_j} 
 (\hat{\Omega}^{-1})_{ij} \, \ln \frac{\zeta_i^*}{\zeta_j}
\label{eq:sea2}
\end{equation}
For consistency with axial current conservation, this must be cancelled by the contribution from the discrete bound states \cite{dt-inprep}. The physical bound state spinors are in general a  (orthonormal) superposition of 
the basis bound states (\ref{eq:bound1}),
$\hat{\psi}^{(i)} = \sum_j C_{ij} \psi^{(j)}
$, where we find that the matrix $C$ is directly related to the matrix $\hat\Omega$ as: $2C\,\hat\Omega C^\dagger ={\bf 1}$.
The density from the bound states (with occupation fractions $\nu_k$)  is
\begin{eqnarray}
\rho_b  
= \sum_{i,j} \frac{{\zeta}_i^*  \zeta_j{\cal N}_1^{(i)*}{\cal N}_1^{(j)} +
{\cal N}_2^{(i)*}{\cal N}_2^{(j)}}{{\cal D}^2\, V_i^*\,V_j}  \sum_k \nu_k C_{ki}^* C_{kj}
\label{eq:bound-rho}
\end{eqnarray}
Then the condition $\rho_{\rm ind}+\rho_b=0$ leads to a consistency condition for the filling 
fractions $\nu_k$ which can be written
\begin{equation}
\nu_k=\frac{1}{2\pi}\,{\rm eigenvalues\,\, of}\left( C^{\dagger\, -1}M \hat{\Omega}^{-1} C^{-1} +h.c.\right)
\label{eq:filling}
\end{equation}
where $M$ is the diagonal matrix $M_{ij} = - i \delta_{i,j}\ln ( -{\zeta}_j^*)$.
Having found a  candidate solution with vanishing fermion density, we now consider the TDHF self-consistency
conditions in (\ref{eq:njl-tdhf}). Eqs.~(\ref{eq:ansatz},\ref{eq:poles}) imply the  
condensate expectation value: 
\begin{eqnarray}
 \langle \bar{\psi}\psi \rangle - i \langle \bar{\psi} i \gamma_5 \psi \rangle & = &  
- \frac{\Delta}{\pi} \ln \Lambda \nonumber \\
& &  \hskip -3cm   - \frac{i}{2\pi} \sum_{i,j} 
\frac{\zeta_i^*{\cal N}_1^{(i)*}}{V_i^* {\cal D}} \frac{{\cal N}_2^{(j)}}{V_j {\cal D}} \, (\hat{\Omega}^{-1})_{ij} \, \ln \frac{\zeta_i^*}{\zeta_j} 
\label{eq:condensate}
\end{eqnarray}
The sum rules satisfied by the residues ${\cal N}_{1,2}^{(i)}$ guarantee UV and IR convergence of the latter terms, while the first term gives self-consistency from the vacuum gap equation
$\frac{Ng^2}{\pi} \ln \Lambda = 1
$.
The second term must be cancelled against the bound state contribution,
and remarkably this is satisfied provided the previously found  filling-fraction condition (\ref{eq:filling}) holds. This proves full TDHF self-consistency for the NJL$_2$ system (\ref{eq:njl-tdhf}). 
For GN$_2$ we impose reality of the condensate $\Delta$ and relax the consistency condition on the pseduoscalar condensate, as discussed below.

We illustrate the TDHF solution (\ref{eq:dets}, \ref{eq:full}, \ref{eq:filling}) with some examples. We write $\zeta_j$ in terms of  phase and boost parameters: 
$\zeta_j=-e^{-i\phi_j}/\eta_j$. With just one pole, $B=e^{2x\sin\phi}$, $A=e^{-2i\phi} b$, and 
$\Delta=(1+e^{-2i\phi}e^{2x\sin\phi} )/(1+e^{2x\sin\phi})$, which is Shei's  twisted kink for the NJL$_2$ model \cite{Shei:1976mn}, with filling fraction $\nu=\phi/\pi$. 
When $\phi=\mp\pi/2$ we get a real solution of the GN$_2$ model, the usual kink/anti-kink.
With two poles we obtain real $\Delta$, for GN$_2$, either by choosing $\phi_1=\phi_2=\pi/2$, which gives
\begin{eqnarray}
\Delta=\frac{1-U_1-U_2+\left(\frac{\eta_1-\eta_2}{\eta_1+\eta_2}\right)^2 U_1 U_2}{1+U_1+U_2+\left(\frac{\eta_1-\eta_2}{\eta_1+\eta_2}\right)^2 U_1 U_2}\,\, , \, U_i\equiv 
\frac{\eta_i|V_i|^2}{2\sin\phi_i}
\label{eq:2kink}
\end{eqnarray}
describing  scattering of 2 kinks, or alternatively by choosing $\zeta_1=-\zeta_2^*$, which means $\phi_2=\pi-\phi_1$ and $\eta_1=\eta_2$ ($=1$ for rest frame). Then $V_2=V_1^*$ and
\begin{eqnarray}
B=
\begin{pmatrix}
{U_1 & \frac{i e^{-i\phi_1}}{2} (V_1^*)^2\cr
-\frac{i e^{i\phi_1}}{2} V_1^2 & U_1}
\end{pmatrix}
, \, A_{ij}=\frac{B_{ij}}{e^{i(\phi_i+\phi_j)}}
\label{2bb}
\end{eqnarray}
Choosing $\omega={\bf 1}$ we obtain the DHN GN$_2$ baryon \cite{Dashen:1974ci}
\begin{eqnarray}
\Delta&=&
\frac{1+2\cos(2\phi_1)U_1+\cos^2(\phi_1) U_1^2}{1+2U_1+\cos^2(\phi_1) U_1^2}\nonumber\\
&=&  1+y\, \tanh\left(y\, \bar x-b\right) - y\, \tanh\left(y\, \bar x+b\right)
\label{eq:dhn-baryon}
\end{eqnarray}
where $y=\sin\phi_1=\tanh(2b)$, and the $x$ origin has been shifted. The GN$_2$ consistency condition leads to filling fractions $\nu_1=\frac{2\phi_1}{\pi}$, $\nu_2=1$. This shows that the DHN GN$_2$ baryon is in fact a bound object of two twisted kinks with filling fractions (which also determine the baryon size) related to the twist angle. Furthermore, the mass of the DHN baryon is related to the masses of the constituent twisted kinks as: $M=M_{\rm kink}(\phi_1)+M_{\rm kink}(\pi-\phi_1)=\frac{2N}{\pi}\sin\phi_1$. 
Choosing instead an off-diagonal mixing matrix 
\begin{eqnarray}
\omega=
\begin{pmatrix}
{\sec\chi & \tan\chi\cr
\tan\chi & \sec\chi}
\end{pmatrix}
\label{eq:off-diagonal}
\end{eqnarray}
leads to the DHN GN$_2$ breather \cite{Dashen:1974ci} 
\begin{widetext}
\begin{eqnarray}
\Delta=\frac{1+2\sec\chi \cos 2\phi_1 U_1\hskip -.1cm -2\tan\chi \sin\phi_1 \cos(2 \bar t \cos\phi_1)U_1+\cos^2\phi_1 U_1^2}{1+2\sec\chi U_1+2\tan\chi \sin\phi_1 \cos(2
 \bar t \cos\phi_1)U_1+\cos^2(\phi_1) U_1^2}\nonumber
\label{eq:dhn-breather}
\end{eqnarray}
\end{widetext}
with filling fractions $\nu_{1,2}=\frac{1}{2\pi}((\phi_1+\phi_2)\pm (\phi_1-\phi_2)\sec\chi)$.
Thus the DHN GN$_2$ breather is  also a bound object of two twisted kinks, with filling fractions related to the twist angles.
\begin{figure}[htb]
\includegraphics[scale=.325]{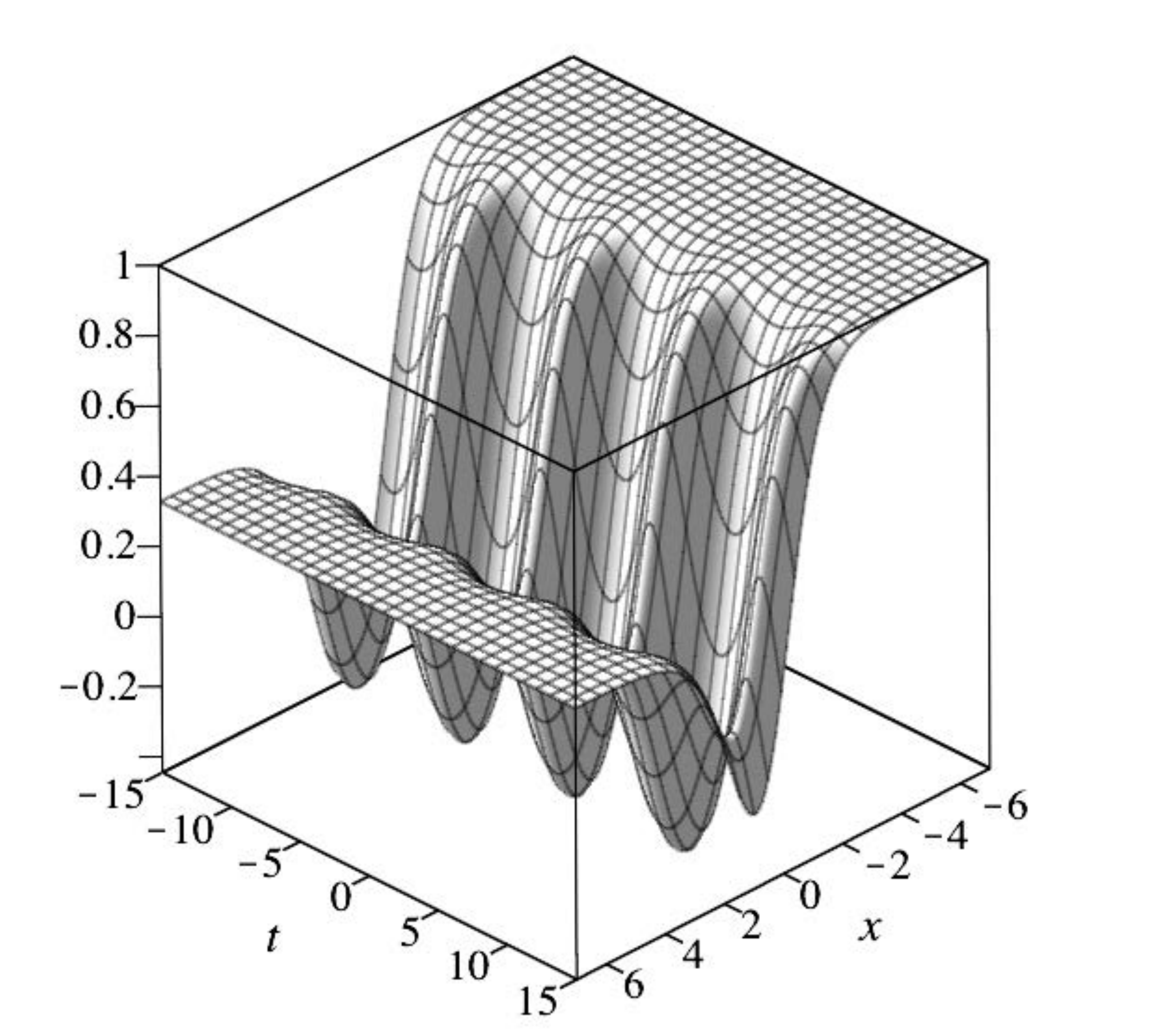}
\includegraphics[scale=.325]{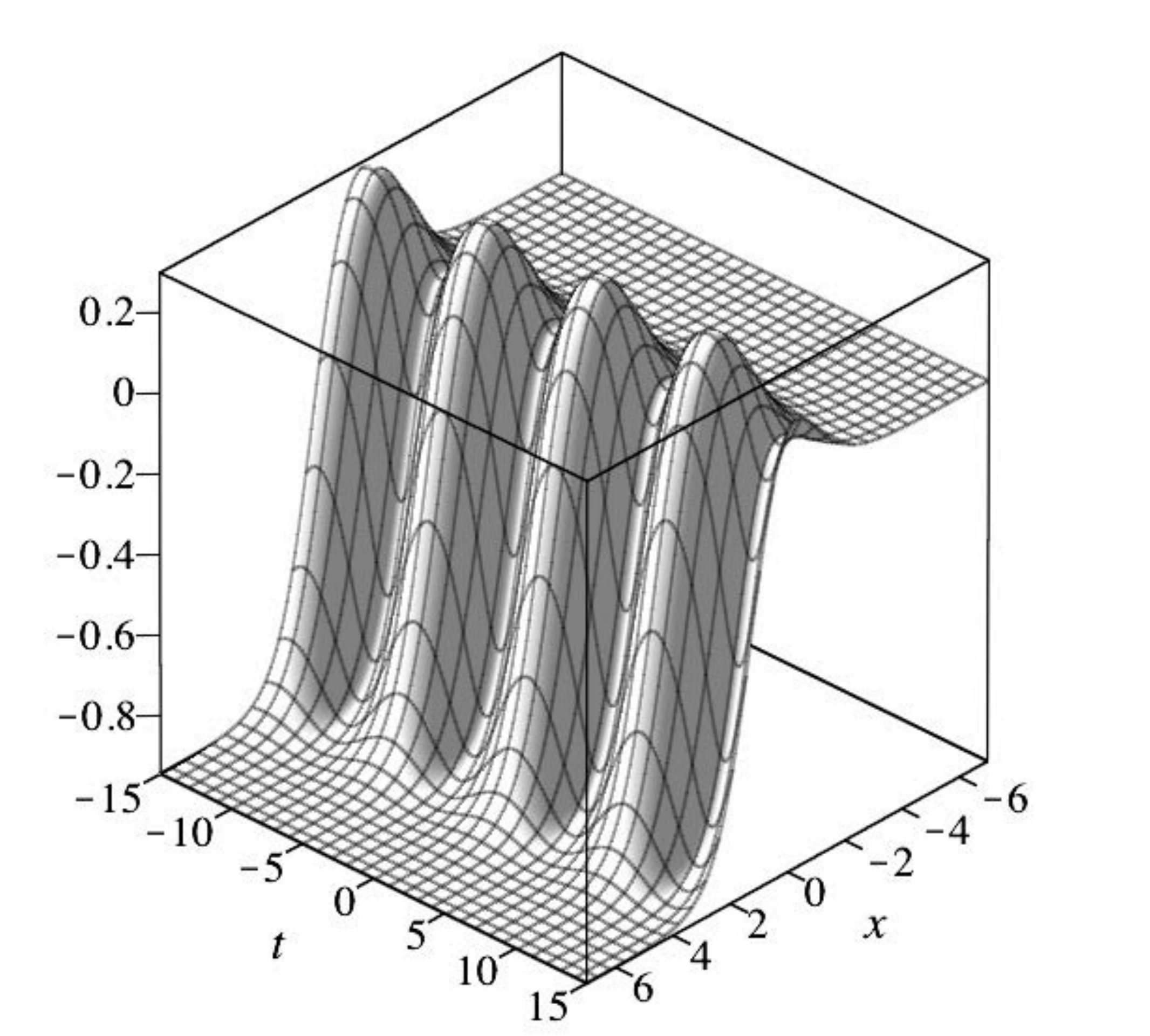}
\caption{The real (upper) and imaginary (lower) parts of the condensate $\Delta(x, t)=S-i P$ for the twisted NJL$_2$ breather. Note the periodic breathing in the time direction of both scalar and pseudoscalar components.}
\label{fig1}
\end{figure}
\begin{figure}[htb]
\includegraphics[scale=.325]{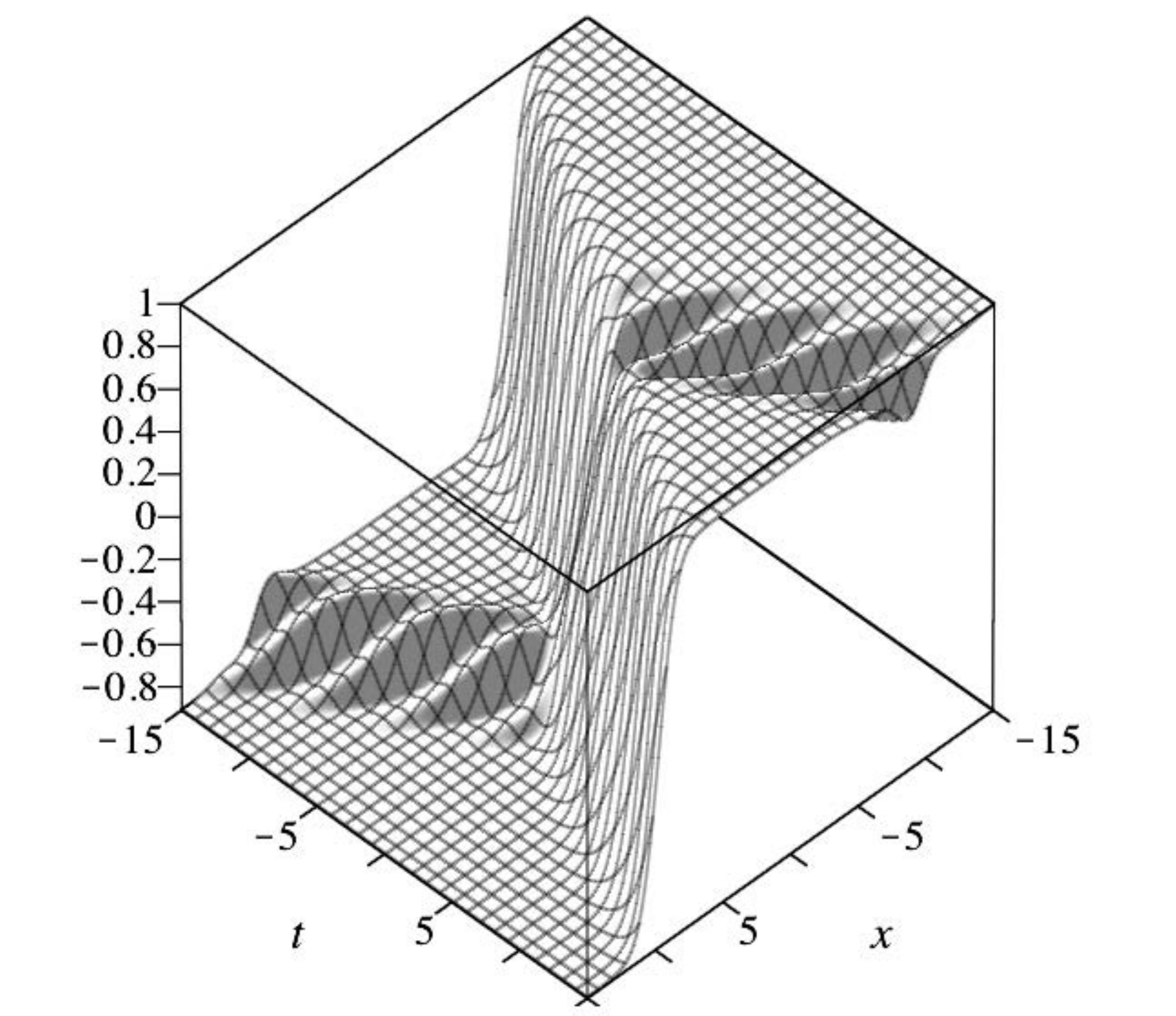}
\includegraphics[scale=.325]{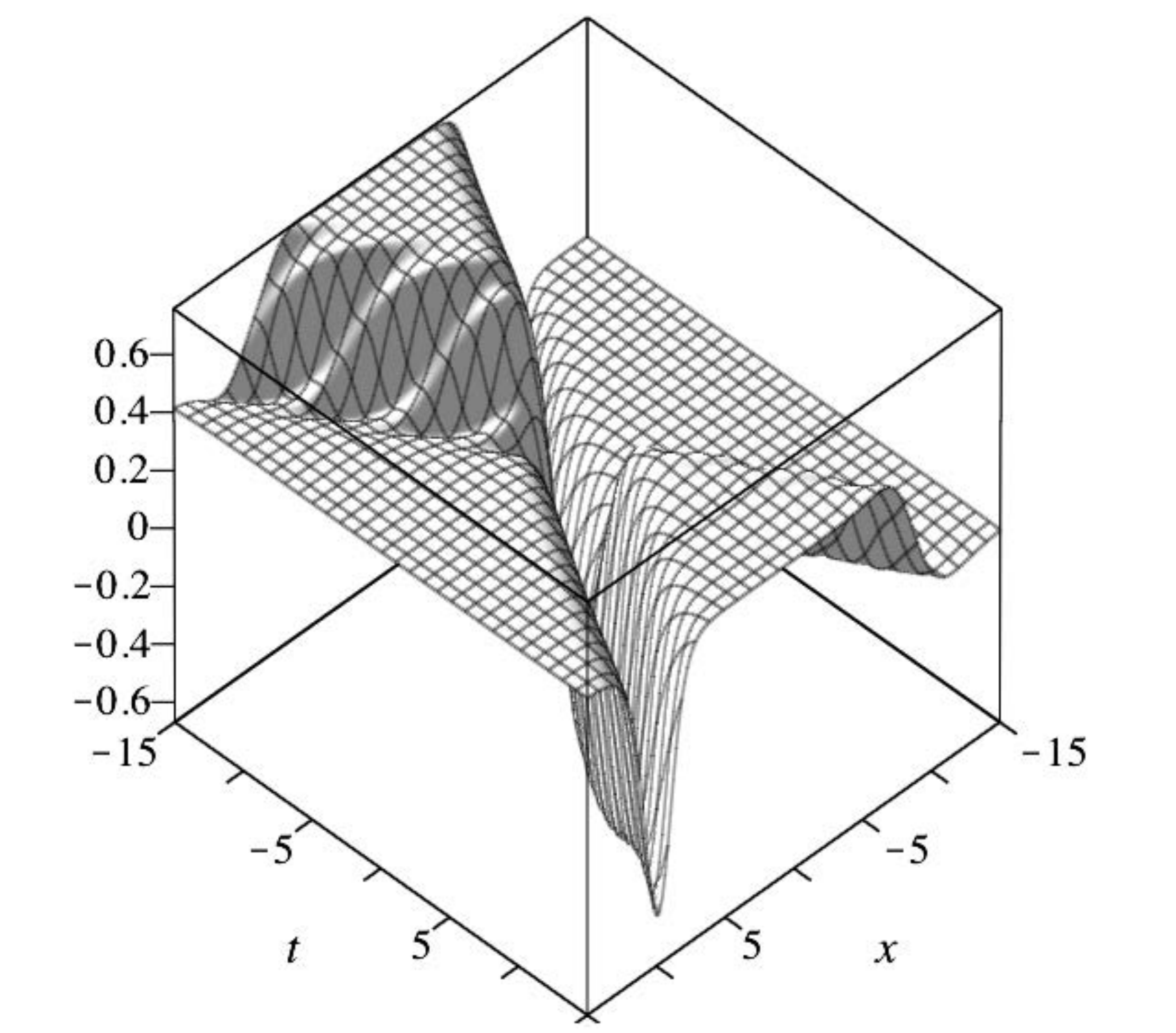}
\caption{The real (upper) and imaginary (lower) parts of the condensate $\Delta(x, t)=S-i P$ for the scattering of two twisted NJL$_2$ breathers. The objects breathe as they scatter.}
\label{fig2}
\end{figure}

Using the off-diagonal form (\ref{eq:off-diagonal}) for $\omega$, but not imposing the reality condition, $\phi_1+\phi_2=\pi$, we obtain a new twisted breather solution to the NJL$_2$ model, 
shown in Fig. \ref{fig1}.
At the three pole level, we find the scattering of three  kinks  if $\omega={\bf 1}$. These are GN$_2$ kinks if $\phi_1=\phi_2=\phi_3=\pi/2$, and twisted NJL$_2$ kinks otherwise. 
The scattering of a GN$_2$ baryon and a kink is obtained by choosing $\zeta_1=-\zeta_2^*$, and the scattering of a breather with a kink is obtained by choosing  the off-diagonal 
form (\ref{eq:off-diagonal}) for a $2\times 2$ sub-block of $\omega$. New breather solutions are obtained by choosing $\eta_1=\eta_2=\eta_3$, and a more general $3\times 3$ 
off-diagonal form of $\omega$.
At the four pole level, in addition to the scattering of 4 (in general twisted) kinks, we can combine the spectral parameters of the kinks pairwise, {\it e.g.} as $\zeta_1=-\zeta_2^*$
 and $\zeta_3=-\zeta_4^*$, to obtain the scattering of two baryons \cite{Dunne:2011wu}. 
 Further choosing the corresponding $2\times 2$ sub-blocks of the mixing matrix $\omega$ to have the breather form
 (\ref{eq:off-diagonal}), we obtain the scattering of two GN$_2$ breathers \cite{Fitzner:2012kb}. Relaxing the pairwise reality conditions, $\phi_1+\phi_2=\pi$ and $\phi_3+\phi_4=\pi$, we obtain another new 
solution,  the scattering of two twisted NJL$_2$ breathers, as shown in Fig. \ref{fig2}. Choosing equal boost parameters $\eta_i$, and a more general off-diagonal mixing matrix $\omega$, we obtain a novel 
4-breather solution in which all 4 twisted kink constituent kinks ``breathe''. 
The pattern should now be clear. Choosing different boost parameters 
$\eta_i$ gives a solution describing scattering  of twisted kink constituents. If some of the $\eta_i$ are equal, the solution describes bound combinations of twisted kink constituents, 
which are baryons if $\omega$ is diagonal, and breathers if  $\omega$ is off-diagonal. The fermion filling fraction consistency condition (\ref{eq:filling}) can always be solved for 
any given choice of spectral parameters $\zeta_i$ and mixing matrix $\omega$. 

In the special case  where all $\phi_i=\frac{\pi}{2}$, the $V_i$ are real, and we have $B_{ij}=\eta_i\eta_j V_i V_j/(\eta_i+\eta_j)$, and $A_{ij}=-B_{ij}$, which agrees with the known 
multi-kink scattering solutions of GN$_2$ \cite{Klotzek:2010gp}, and whose non-relativistic limit agrees with the Schr\"odinger results of Kay-Moses and Nogami-Warke 
\cite{kaymoses,nogami}. For NJL$_2$, taking all $\eta_i=1$ we obtain the static solution $\Delta=\det({\bf 1}+\hat A)/\det({\bf 1}+\hat B)$, with
\begin{eqnarray}
\hat B_{ij}=\frac{e^{(x-x_i)\sin\phi_i+(x-x_j)\sin\phi_j}}{2\sin\left(\frac{\phi_i+\phi_j}{2}\right)}\, , \, \hat A_{ij}=\frac{\hat  B_{ij}}{e^{i(\phi_i+\phi_j)}}
\label{eq:nitta}
\end{eqnarray}
where we have removed phases from the determinant, which is possible for diagonal $\omega$. This is a compact solution for the algebraic system found recently in 
\cite{Takahashi:2012pk} for self-consistent {\it static} multi-twisted kinks in NJL$_2$. Taking the $\eta_i$ different, our solution (\ref{eq:dets}, \ref{eq:full}, \ref{eq:filling}) gives the full 
 time-dependent generalization.
\\

We acknowledge support  from DOE grant DE-FG02-92ER40716 (GD) and  DFG grant TH 842/1-1 (MT).


\begin{thebibliography}{999}

\bibitem{Nambu:1961tp} 
  Y.~Nambu, G.~Jona-Lasinio,
  ``Dynamical Model of Elementary Particles Based on an Analogy with Superconductivity. 1.,''
  Phys.\ Rev.\  {\bf 122}, 345 (1961).
  
\bibitem{Fulde:1964zz} 
  P.~Fulde and R.~A.~Ferrell,
  ``Superconductivity in a Strong Spin-Exchange Field,''
  Phys.\ Rev.\  {\bf 135}, A550 (1964).
  A.~I.~Larkin and Y.~N.~Ovchinnikov,
  ``Nonuniform state of superconductors,''
  Zh.\ Eksp.\ Teor.\ Fiz.\  {\bf 47}, 1136 (1964)
  [Sov.\ Phys.\ JETP {\bf 20}, 762 (1965)].
  
      \bibitem{peierls}
R.~Peierls, {\it The Quantum Theory of Solids} (Oxford, 1955);
P.~G.~de~Gennes, {\it Superconductivity of Metals and Alloys} (Addison-Wesley, Redwood City, CA, 1989).



  \bibitem{rajagopal}
  K.~Rajagopal and F.~Wilczek,
  ``The condensed matter physics of QCD,'' 
  in {\it At the Frontier of Particle Physics; Handbook of QCD}, 
  M. Shifman, ed., (World Scientific, 2001),
\hhref{hep-ph/0011333}.
  
   \bibitem{casalbuoni}
  R.~Casalbuoni and G.~Nardulli,
  ``Inhomogeneous superconductivity in condensed matter and QCD,''
  Rev.\ Mod.\ Phys.\  {\bf 76}, 263 (2004), 
\hhref{hep-ph/0305069}.
  
\bibitem{Heeger:1988zz} 
  A.~J.~Heeger, S.~Kivelson, J.~R.~Schrieffer and W.~-P.~Su,
  ``Solitons in conducting polymers,''
  Rev.\ Mod.\ Phys.\  {\bf 60}, 781 (1988).
  
 
 
   \bibitem{zwierlein} M.~W.~Zwierlein, A.~Schirotzek, C.~H.~Schunck and W.~Ketterle, ``Fermionic Superfluidity with Imbalanced Spin Populations and the Quantum Phase Transition to the Normal State'',
Science {\bf 311}, 492 (2006);
G.~B.~Partridge {\it et al}, 
 ``Pairing and Phase Separation in a Polarized Fermi Gas'', 
Science {\bf 311}, 503 (2006).
  
\bibitem{pitaevskii} 
  S.~Giorgini, L.~P.~Pitaevskii and S.~Stringari,
  ``Theory of ultracold atomic Fermi gases,''
  Rev.\ Mod.\ Phys.\  {\bf 80}, 1215 (2008), 
\hhref{0706.3360}.
  
\bibitem{Adams:2012th} 
  A.~Adams, L.~D.~Carr, T.~Sch\"afer, P.~Steinberg and J.~E.~Thomas,
  ``Strongly Correlated Quantum Fluids: Ultracold Quantum Gases, Quantum Chromodynamic Plasmas, and Holographic Duality,''
  New J.\ Phys.\  {\bf 14}, 115009 (2012),
\hhref{1205.5180}.



  
\bibitem{Herzog:2007ij} 
  C.~P.~Herzog, P.~Kovtun, S.~Sachdev and D.~T.~Son,
  ``Quantum critical transport, duality, and M-theory,''
  Phys.\ Rev.\ D {\bf 75}, 085020 (2007),
  \hhref{hep-th/0701036};
  S.~Sachdev,
  ``Holographic metals and the fractionalized Fermi liquid,''
  Phys.\ Rev.\ Lett.\  {\bf 105}, 151602 (2010),
\hhref{1006.3794}.

  

    \bibitem{jackiw}
  R. Jackiw and C. Rebbi, 
    ``Solitons with Fermion Number 1/2,''
  Phys.\ Rev.\ D {\bf 13}, 3398 (1976);
  A.~J.~Niemi and G.~W.~Semenoff,
  ``Fermion Number Fractionization in Quantum Field Theory,''
  Phys.\ Rept.\  {\bf 135}, 99 (1986).
 

\bibitem{Gross:1974jv} 
  D.~J.~Gross and A.~Neveu,
  ``Dynamical Symmetry Breaking in Asymptotically Free Field Theories,''
  Phys.\ Rev.\ D {\bf 10}, 3235 (1974).
  
\bibitem{Thies:2006ti} 
  M.~Thies,
  ``From relativistic quantum fields to condensed matter and back again: Updating the Gross-Neveu phase diagram,''
  J.\ Phys.\ A {\bf 39}, 12707 (2006),
  \hhref{hep-th/0601049}.
  
\bibitem{Dashen:1974ci} 
  R.~F.~Dashen, B.~Hasslacher and A.~Neveu,
  ``Semiclassical Bound States in an Asymptotically Free Theory,''
  Phys.\ Rev.\ D {\bf 12}, 2443 (1975).
  
\bibitem{Feinberg:2003qz} 
  J.~Feinberg,
  ``All about the static fermion bags in the Gross-Neveu model,''
  Annals Phys.\  {\bf 309}, 166 (2004),
  \hhref{hep-th/0305240}.

  
\bibitem{Basar:2009fg} 
  G.~Basar, G.~V.~Dunne and M.~Thies,
  ``Inhomogeneous Condensates in the Thermodynamics of the Chiral NJL(2) model,''
  Phys.\ Rev.\ D {\bf 79}, 105012 (2009),
\hhref{0903.1868}.
  

  
  \bibitem{horovitz}
B.~Horovitz,
``Soliton Lattice in Polyacetylene, Spin-Peierls Systems, and Two-Dimensional Sine-Gordon 
 Systems'',
Phys.\ Rev.\ Lett.\ {\bf 46}, 742  (1981).

\bibitem{braz}
S.~A.~Brazovskii and N.~N.~Kirova, 
``Excitons, polarons and bipolarons in conducting polymers'', Pis. Zh. Eksp. Teor. Fiz. {\bf 33}, 6 (1981) 
[JETP Lett. {\bf 33}, 4 (1981)].


\bibitem{machida}
K.~Machida and H.~Nakanishi,
``Superconductivity under a ferromagnetic molecular field'', 
Phys.\ Rev.\ B {\bf 30}, 122  (1984).

\bibitem{Takahashi:2012pk} 
  D.~A.~Takahashi and M.~Nitta,
  ``Self-consistent multiple complex-kink solutions in Bogoliubov-de Gennes and chiral Gross-Neveu systems,''
  Phys.\  Rev.\  Lett.\  {\bf 110},  131601 (2013),
\hhref{1209.6206}.
  
   \bibitem{integrable}
  A.~Neveu and N.~Papanicolaou,
  ``Integrability of the Classical Scalar and Symmetric Scalar-Pseudoscalar Contact Fermi Interactions in Two-Dimensions,''
  Commun.\ Math.\ Phys.\  {\bf 58}, 31 (1978);
  V.~E.~Zakharov and A.~V.~Mikhailov,
  ``On The Integrability Of Classical Spinor Models In Two-dimensional Space-time,''
  Commun.\ Math.\ Phys.\  {\bf 74}, 21 (1980).

\bibitem{Shei:1976mn} 
  S.~-S.~Shei,
  ``Semiclassical Bound States in a Model with Chiral Symmetry,''
  Phys.\ Rev.\ D {\bf 14}, 535 (1976).
  
\bibitem{Dunne:2011wu} 
  G.~V.~Dunne, C.~Fitzner and M.~Thies,
  ``Baryon-baryon scattering in the Gross-Neveu model: the large N solution,''
  Phys.\ Rev.\ D {\bf 84}, 105014 (2011),
\hhref{1108.5888};
  C.~Fitzner and M.~Thies,
  ``Evidence for factorized scattering of composite states in the Gross-Neveu model,''
  Phys.\ Rev.\ D {\bf 85}, 105015 (2012),
\hhref{1202.0648}.

\bibitem{Klotzek:2010gp} 
  A.~Klotzek and M.~Thies,
  ``Kink dynamics, sinh-Gordon solitons and strings in AdS$_3$ from the Gross-Neveu model,''
  J.\ Phys.\ A {\bf 43}, 375401 (2010),
\hhref{1006.0324};
  C.~Fitzner and M.~Thies,
  ``Exact solution of N baryon problem in the Gross-Neveu model,''
  Phys.\ Rev.\ D {\bf 83}, 085001 (2011),
\hhref{1010.5322}.
  
\bibitem{Jevicki:2009uz} 
  A.~Jevicki and K.~Jin,
 ``Moduli Dynamics of AdS(3) Strings,''
  JHEP {\bf 0906}, 064 (2009),
\hhref{0903.3389}.
  
    \bibitem{dt-inprep}
   G.~V.~Dunne and M.~Thies, 
   ``Full Time-Dependent Hartree-Fock Solution of Large N Gross-Neveu models'', to appear.

\bibitem{kaymoses}
I. Kay and H. E. Moses, 
``Reflectionless transmission through dielectrics and scattering potentials'',
J. Appl. Phys. {\bf 27}, 1503 (1956).

\bibitem{nogami}
Y. Nogami and C. Warke, 
``Soliton solutions of multicomponent nonlinear Schršdinger equation'',
Phys. Lett. A {\bf 59}, 251 (1976).

\bibitem{Fitzner:2012kb} 
  C.~Fitzner and M.~Thies,
  ``Breathers and their interaction in the massless Gross-Neveu model,''
  Phys.\ Rev.\ D {\bf 87}, 025001 (2013),
\hhref{1210.4423}.
  
  

 

\end{thebibliography}
\end{document}